\documentclass[11pt]{article}
\usepackage[utf8]{inputenc}
\usepackage[T1]{fontenc}
\usepackage{amsmath,amssymb}
\usepackage{graphicx}
\usepackage{booktabs}
\usepackage{geometry}
\usepackage{microtype}
\usepackage{enumitem}
\usepackage[hidelinks]{hyperref}
\title{\textbf{Fisher-Information Design of Ridge-Loaded Subwavelength Slits}}
\author{
Juan Sumaya-Mart\'inez$^{1,*}$ \and Omar Olmos-L\'opez$^{2}$\\[0.4em]
\small $^{1}$Faculty of Sciences, Universidad Aut\'onoma del Estado de M\'exico,\\
\small Toluca 50000, Estado de M\'exico, Mexico\\
\small $^{2}$School of Engineering and Sciences, Tecnol\'ogico de Monterrey,\\
\small Toluca, Estado de M\'exico, Mexico\\[0.3em]
\small $^*$Corresponding author: \texttt{jsm@uaemex.mx}\\
\small ORCID: J.S.-M. 0000-0002-7032-8824; O.O.-L. 0000-0002-3431-1160
}
\date{August 2026}

\begin{document}
\maketitle

\begin{abstract}
We present a compact information-theoretic description of extraordinary optical transmission through a subwavelength slit containing symmetric internal ridges. Full-wave finite-element spectra are interpreted with a reduced Fabry--P\'erot model in which the ridge interfaces contribute a reactive discontinuity phase. Fisher information is then used to quantify how reliably resonance position and geometric parameters can be inferred from transmission data. The framework clarifies three points that are easily missed by peak-based optimization: a ridge-induced spectral shift is controlled by both propagation and interface phase; a negative effective cavity correction does not by itself imply a larger phase velocity; and the resonance with the highest quality factor is not necessarily the most informative under realistic noise. For an isolated Lorentzian resonance under white Gaussian noise, the Fisher information for the resonance center scales linearly with $Q$ when amplitude, sampling, and noise are fixed. We also give the corresponding forms for correlated Gaussian noise and Poisson counting noise and outline a reduced-order inverse-design workflow. The result is a concise route from electromagnetic response to parameter-estimation precision in ridge-loaded subwavelength resonators.
\end{abstract}

\section{Introduction}
Extraordinary optical transmission (EOT) through subwavelength apertures is a canonical example of resonant light transport in structured metallic systems \cite{Ebbesen1998,Porto1999,Takakura2001,Liu2008,Rodrigo2016}. Depending on the material and geometry, transmission enhancement can involve surface waves, diffraction-driven funneling, and Fabry--P\'erot (FP) resonances of modes confined inside the aperture \cite{Astilean2000,Pardo2011,Li2018}. A perfect-electric-conductor (PEC) slit is particularly useful as a reference system because it isolates geometric and cavity effects from ohmic loss.

Internal metallic ridges add a second length scale to the slit. By locally reducing the aperture width, they alter modal confinement, impedance, and round-trip phase. The usual design question is where a resonance occurs or how large its transmission becomes. For sensing and inference, however, the more relevant question is how strongly the measured spectrum changes when a parameter changes and whether that change remains distinguishable in noise.

Here we formulate that question with Fisher information (FI). The electromagnetic model provides the spectrum, a reduced FP description explains the dominant phase shifts, and FI converts the spectral derivative into an estimation metric. The purpose of this short preprint is not to introduce a new EOT mechanism, but to show how a familiar resonant structure can be ranked by \emph{identifiability} rather than by transmission alone.

\section{Geometry and electromagnetic model}
\begin{figure}[ht]
\centering
\includegraphics[width=0.70\textwidth]{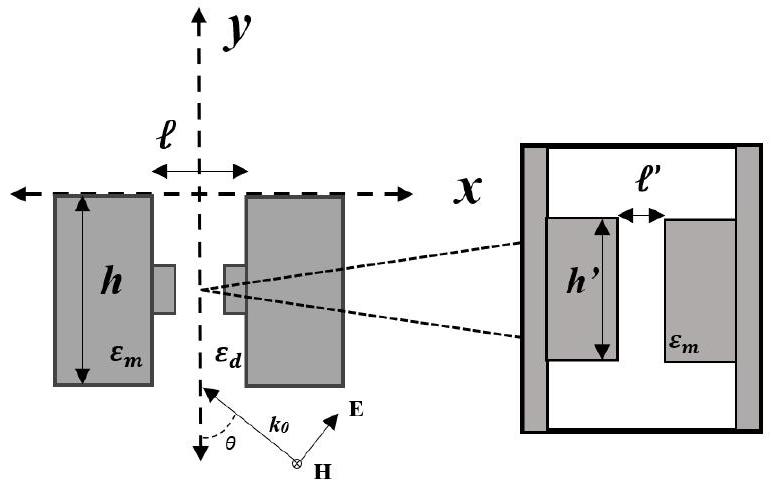}
\caption{Ridge-loaded subwavelength slit used in the full-wave calculations.}
\label{fig:geometry}
\end{figure}

We consider a two-dimensional slit of width $\ell$ in a metallic film of thickness $h$. Two symmetric internal ridges reduce the local gap to $\ell'$ over a vertical extent $h'$, as shown in Fig.~\ref{fig:geometry}. The incident field is TM polarized, with magnetic field along the invariant $z$ direction.

With the time convention $\exp(-i\omega t)$ and $\mathbf H=H_z(x,y)\hat{\mathbf z}$, the scalar TM equation in an isotropic medium is
\begin{equation}
\nabla\!\cdot\!\left(\varepsilon^{-1}\nabla H_z\right)+\omega^2\mu H_z=0.
\label{eq:tm}
\end{equation}
On PEC boundaries, the tangential electric field vanishes, which in the $H_z$ formulation gives $\partial_nH_z=0$. Open boundaries are terminated with perfectly matched layers. The electric field is reconstructed from
\begin{equation}
\mathbf E=\frac{i}{\omega\varepsilon}\nabla\times(H_z\hat{\mathbf z}),
\end{equation}
and the time-averaged energy flux is
\begin{equation}
\mathbf S=\frac12\operatorname{Re}\{\mathbf E\times\mathbf H^*\}.
\end{equation}
The transmitted power is obtained by integrating the normal component of $\mathbf S$ across the output plane and normalizing to the incident flux.

The calculations summarized below use $\ell=1~\mu$m and $h=20~\mu$m, with wavelengths between 10 and 45~$\mu$m. The PEC approximation is used to isolate the geometric phase physics. Real-metal loss is discussed in Sec.~\ref{sec:loss}.

\section{Reduced Fabry--P\'erot description}
For a uniform slit, the $m$th transmission maximum is represented by
\begin{equation}
\lambda_m \simeq \frac{2n_{\rm eff}}{m}\left(h+\Delta_{\rm end}\right),
\label{eq:base}
\end{equation}
where $\Delta_{\rm end}$ is an entrance/exit phase correction. In the present PEC regime, the fundamental mode has $n_{\rm eff}\simeq1$.

With internal ridges, the slit is treated as two serial sections with propagation constants $\beta_1$ and $\beta_2$ and modal impedances $Z_1$ and $Z_2$. The resonance condition becomes
\begin{equation}
\beta_1(h-h')+\beta_2h'+\phi_{\rm in}+\phi_{\rm out}+\phi_{\rm disc}=m\pi.
\label{eq:phase}
\end{equation}
A compact single-mode approximation for the discontinuity contribution is
\begin{equation}
\phi_{\rm disc}\simeq2\arctan\!\left[
\frac{Z_2-Z_1}{Z_2+Z_1}\tan(\beta_2h')\right],
\label{eq:disc}
\end{equation}
with
\begin{equation}
Z_j=\frac{\langle E_t\rangle_j}{\langle H_z\rangle_j}.
\end{equation}

The ratio $(Z_2-Z_1)/(Z_2+Z_1)$ is the reflection coefficient of the retained fundamental channel. At an abrupt step, higher-order evanescent modes are also excited. Eliminating those localized degrees of freedom leaves an effective reactive loading of the propagating mode, which is represented by $\phi_{\rm disc}$. A full mode-matching calculation would explicitly solve for those evanescent amplitudes and recover the interface phase from vectorial boundary continuity.

It is convenient to rewrite Eq.~(\ref{eq:phase}) as
\begin{equation}
\lambda_m\simeq\frac{2n_{\rm eff}}{m}
\left[h+\Delta_{\rm end}+\Delta_{\rm ridge}\right].
\label{eq:effective}
\end{equation}
The quantity $\Delta_{\rm ridge}$ is an \emph{effective phase correction}. Its sign does not directly determine the local phase velocity. A negative value can arise from the reflection phase at the two interfaces even if the local propagation constant in the narrow section has not decreased.

\section{Representative full-wave results}
\begin{figure}[ht]
\centering
\includegraphics[width=0.92\textwidth]{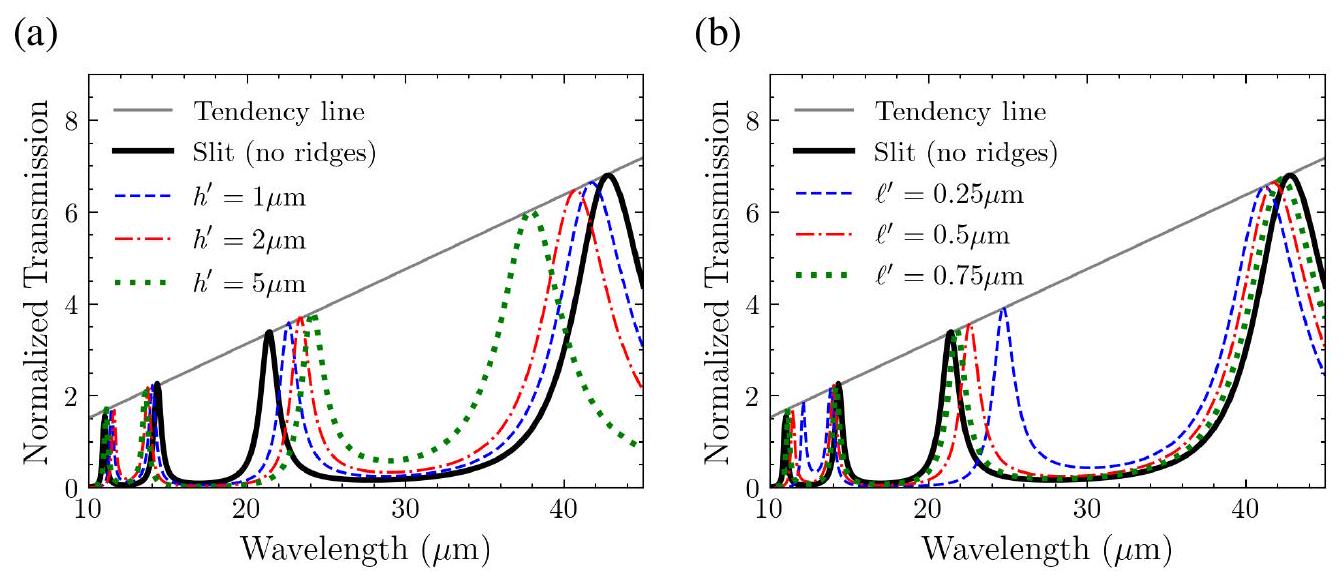}
\caption{Transmission spectra for ridge-loaded slits. Left: fixed $\ell'=0.5~\mu$m and varying $h'$. Right: fixed $h'=1~\mu$m and varying $\ell'$.}
\label{fig:spectra}
\end{figure}

Figure~\ref{fig:spectra} shows representative ridge sweeps. Varying $h'$ or $\ell'$ moves different resonances by different amounts and can shift different modes in opposite directions. This behavior is consistent with a mode-dependent discontinuity phase rather than a single universal geometric scaling.

The resonance positions used in the reduced model are summarized in Table~\ref{tab:res}. A calibration of the plain slit gives $\Delta_{\rm end}\simeq1.55~\mu$m. For moderate ridge perturbations, a single fitted $\Delta_{\rm ridge}$ provides a useful first approximation; for deep or very narrow ridge gaps, the mode dependence of Eq.~(\ref{eq:disc}) becomes important.

\begin{table}[ht]
\centering
\caption{Selected resonance positions in micrometers.}
\label{tab:res}
\begin{tabular}{ccccc}
\toprule
Configuration & $\lambda_1$ & $\lambda_2$ & $\lambda_3$ & $\lambda_4$\\
\midrule
No ridges & 11.0 & 14.3 & 21.4 & 42.7\\
$h'=1~\mu$m, $\ell'=0.5~\mu$m & 11.4 & 14.0 & 22.6 & 41.7\\
$h'=2~\mu$m, $\ell'=0.5~\mu$m & 11.6 & 13.7 & 23.3 & 40.7\\
$h'=1~\mu$m, $\ell'=0.25~\mu$m & 12.1 & 13.8 & 24.7 & 41.1\\
\bottomrule
\end{tabular}
\end{table}

\begin{figure}[ht]
\centering
\includegraphics[width=0.88\textwidth]{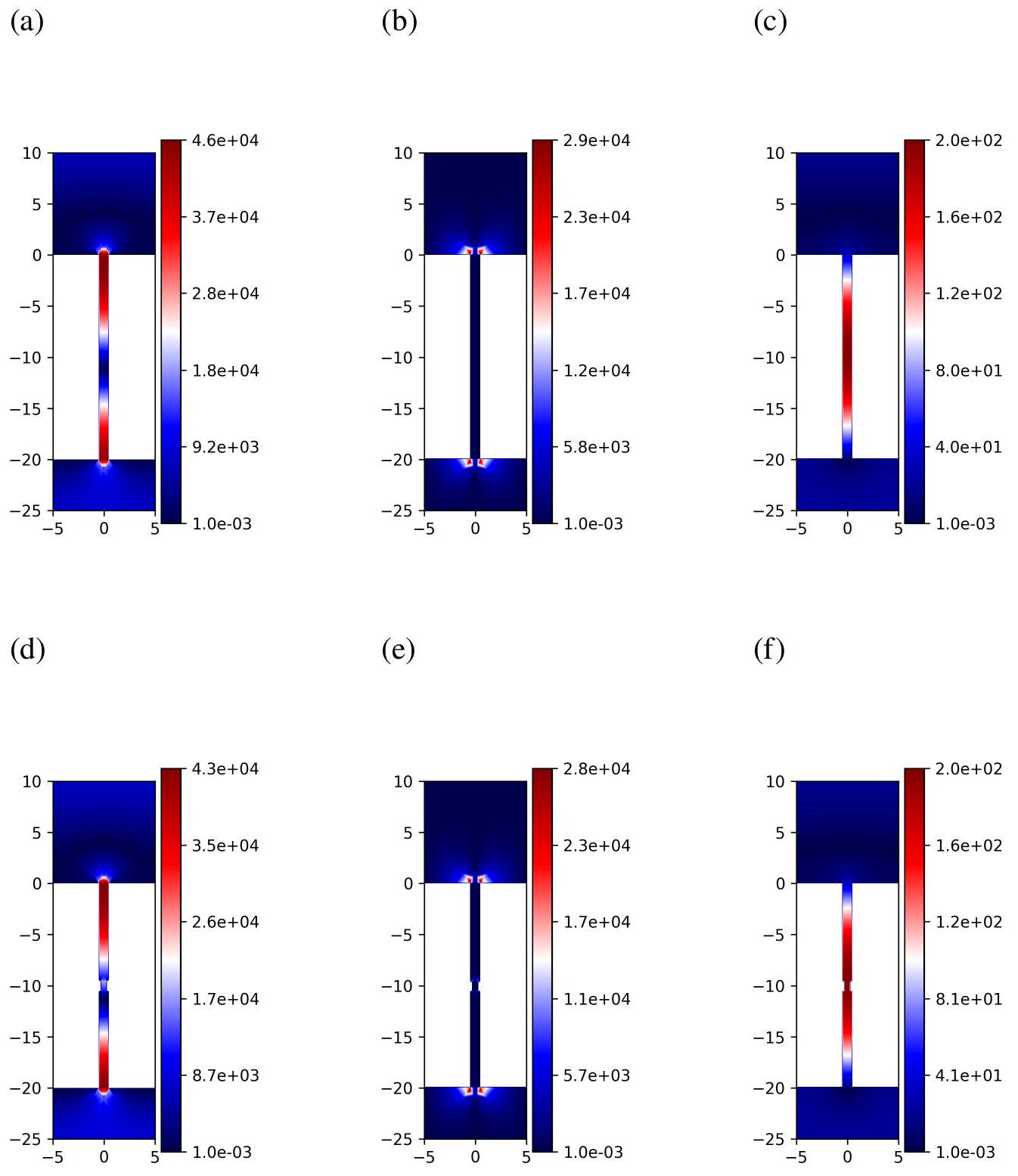}
\caption{Representative resonant field distributions for a plain slit and a ridge-loaded slit.}
\label{fig:fields}
\end{figure}

The near fields in Fig.~\ref{fig:fields} illustrate the same physics in real space. The internal interfaces introduce additional localized field structure while modifying the standing-wave pattern. In the PEC model these hot spots represent reactive field concentration and redistribution of energy flow, not ohmic heating.

\begin{figure}[ht]
\centering
\includegraphics[width=0.62\textwidth]{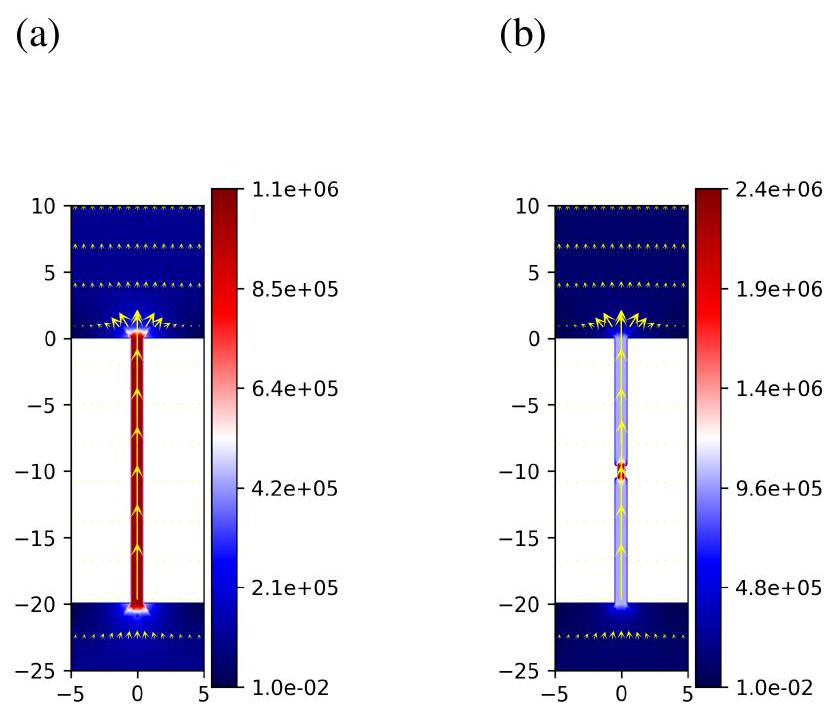}
\caption{Poynting-vector magnitude and direction at resonance for the plain and ridge-loaded configurations.}
\label{fig:poynting}
\end{figure}

The Poynting-vector map in Fig.~\ref{fig:poynting} shows stronger localized energy flow in the resonant ridge-loaded case.

\section{Fisher information for spectral inference}
Let the measured spectrum be a vector $\mathbf y$ with mean $\boldsymbol\mu(\boldsymbol\theta)$, where $\boldsymbol\theta$ contains the parameters to be estimated. The Fisher-information matrix is
\begin{equation}
F_{ab}=\mathbb E\!\left[
\frac{\partial\ln p}{\partial\theta_a}
\frac{\partial\ln p}{\partial\theta_b}\right].
\end{equation}
For additive independent Gaussian noise of variance $\sigma^2$,
\begin{equation}
F_{ab}=\frac{1}{\sigma^2}\sum_i
\frac{\partial\mu_i}{\partial\theta_a}
\frac{\partial\mu_i}{\partial\theta_b}.
\label{eq:fiwhite}
\end{equation}
The Cram\'er--Rao bound gives
\begin{equation}
\operatorname{Cov}(\hat{\boldsymbol\theta})\succeq\mathbf F^{-1}.
\end{equation}
Thus a geometry is informative when small parameter changes generate large, nondegenerate changes in the detected spectrum.

For correlated Gaussian noise with covariance $\mathbf C$ independent of the parameters,
\begin{equation}
F_{ab}=
\boldsymbol\mu_{,a}^{T}\mathbf C^{-1}\boldsymbol\mu_{,b}.
\label{eq:correlated}
\end{equation}
Correlated source fluctuation or detector drift therefore down-weights spectral directions that overlap strongly with noisy covariance modes.

For photon counting with independent Poisson means $\mu_i$,
\begin{equation}
F_{ab}^{(P)}=\sum_i\frac{1}{\mu_i}
\frac{\partial\mu_i}{\partial\theta_a}
\frac{\partial\mu_i}{\partial\theta_b}.
\label{eq:poisson}
\end{equation}
This weighting makes the optimum depend on detected photon level as well as spectral slope.

\subsection{Relation between FI and resonance quality factor}
For an isolated Lorentzian
\begin{equation}
T(\lambda)=B+\frac{A}{1+4(\lambda-\lambda_0)^2/\Gamma^2},
\qquad Q=\frac{\lambda_0}{\Gamma},
\end{equation}
sampled uniformly with spacing $\Delta\lambda$ under white Gaussian noise, the FI for the center wavelength is
\begin{equation}
F_{\lambda_0}\simeq
\frac{\pi A^2}{2\sigma^2\Delta\lambda\,\Gamma}
=
\frac{\pi A^2}{2\sigma^2\Delta\lambda}\frac{Q}{\lambda_0}.
\label{eq:fiq}
\end{equation}
Under these restrictive conditions, FI grows linearly with $Q$. This does \emph{not} make the highest-$Q$ resonance universally optimal. A narrow resonance can be undersampled, have low contrast, overlap with another mode, or align with a drifting baseline. Under Poisson noise, the signal-dependent weighting in Eq.~(\ref{eq:poisson}) can also shift the optimum. FI therefore provides a more complete design metric than $Q$ alone.

\section{Implications for design}
\subsection{Dispersive metallic loss}
\label{sec:loss}
For a real metal, a Drude model
\begin{equation}
\varepsilon_m(\omega)=\varepsilon_\infty-
\frac{\omega_p^2}{\omega^2+i\gamma\omega}
\end{equation}
makes both $\beta$ and $Z$ complex. The resonance then acquires absorptive as well as radiative broadening,
\begin{equation}
\frac{1}{Q_{\rm loaded}}=
\frac{1}{Q_{\rm rad}}+\frac{1}{Q_{\rm ohm}}.
\end{equation}
Because metal overlap depends on ridge geometry, a PEC optimum is not guaranteed to remain optimal for gold or silver. Loss-aware design must recompute both the transmission spectrum and the corresponding FI.

\subsection{Reduced-order inverse design}
Let $\mathbf g=(\ell',h',\ldots)$ denote the geometry and $n$ a refractive index to be estimated. A practical workflow is:
\begin{enumerate}[leftmargin=1.5em]
\item use the FP/discontinuity model as a fast surrogate $T(\lambda;\mathbf g,n)$;
\item compute the relevant FI matrix under the experimental noise model;
\item minimize $[\mathbf F^{-1}]_{nn}$ subject to fabrication and transmission constraints;
\item validate the best candidates with full-wave FEM;
\item update the surrogate with the high-fidelity results.
\end{enumerate}
This separation between fast exploration and full-wave validation makes FI-based optimization computationally practical.

\section{Conclusions}
A ridge-loaded subwavelength slit can be viewed simultaneously as a resonant electromagnetic system and as an information channel. The reduced FP model explains how propagation and interface phases jointly determine the resonance shifts, while the FI formalism evaluates how precisely those shifts and geometric parameters can be inferred from noisy spectra.

The main result is conceptual but practical: maximizing transmission or $Q$ is not identical to maximizing parameter identifiability. Under ideal white noise, the FI of an isolated Lorentzian center scales as $Q$, but correlated noise, photon statistics, contrast, overlap, and finite sampling can change the optimum. Internal ridges therefore provide not only a way to tune EOT resonances but also a geometry with which to shape the information content of the spectrum.

The same strategy extends naturally to dispersive metals, refractive-index sensing, and more complex resonant apertures, provided that the forward model and the experimental noise statistics are treated consistently.

\section*{Data availability}
The numerical data underlying the representative spectra and resonance positions are available from the corresponding author upon reasonable request.

\section*{Competing interests}
The authors declare no competing interests.


\begin{thebibliography}{99}

\bibitem{Ebbesen1998}
T. W. Ebbesen, H. J. Lezec, H. F. Ghaemi, T. Thio, and P. A. Wolff,
``Extraordinary optical transmission through sub-wavelength hole arrays,''
Nature \textbf{391}, 667--669 (1998).

\bibitem{Porto1999}
J. A. Porto, F. J. Garc\'ia-Vidal, and J. B. Pendry,
``Transmission resonances on metallic gratings with very narrow slits,''
Phys. Rev. Lett. \textbf{83}, 2845--2848 (1999).

\bibitem{Takakura2001}
Y. Takakura,
``Optical resonance in a narrow slit in a thick metallic screen,''
Phys. Rev. Lett. \textbf{86}, 5601--5603 (2001).

\bibitem{Astilean2000}
S. Astilean, P. Lalanne, and M. Palamaru,
``Light transmission through metallic channels much smaller than the wavelength,''
Opt. Commun. \textbf{175}, 265--273 (2000).

\bibitem{Liu2008}
H. T. Liu and P. Lalanne,
``Microscopic theory of the extraordinary optical transmission,''
Nature \textbf{452}, 728--731 (2008).

\bibitem{Pardo2011}
F. Pardo, P. Bouchon, R. Ha\"idar, and J. L. Pelouard,
``Light funneling mechanism explained by magnetoelectric interference,''
Phys. Rev. Lett. \textbf{107}, 093902 (2011).

\bibitem{Li2018}
J. W. Li, J. S. Hong, W. T. Chou, D. J. Huang, and K. R. Chen,
``Light funneling profile during enhanced transmission through a subwavelength metallic slit,''
Plasmonics \textbf{13}, 2249--2254 (2018).

\bibitem{Rodrigo2016}
S. G. Rodrigo, F. de Le\'on-P\'erez, and L. Mart\'in-Moreno,
``Extraordinary optical transmission: Fundamentals and applications,''
Proc. IEEE \textbf{104}, 2288--2306 (2016).

\bibitem{Nikitin2008}
A. Y. Nikitin, D. Zueco, F. J. Garc\'ia-Vidal, and L. Mart\'in-Moreno,
``Electromagnetic wave transmission through a small hole in a perfect electric conductor of finite thickness,''
Phys. Rev. B \textbf{78}, 165429 (2008).

\bibitem{Chang2015a}
S. H. Chang and Y. L. Su,
``Mapping of transmission spectrum between plasmonic and nonplasmonic single slits. I: Resonant transmission,''
J. Opt. Soc. Am. B \textbf{32}, 38--44 (2015).

\bibitem{Chang2015b}
S. H. Chang and Y. L. Su,
``Mapping of transmission spectrum between plasmonic and nonplasmonic single slits. II: Nonresonant transmission,''
J. Opt. Soc. Am. B \textbf{32}, 45--51 (2015).

\bibitem{GarciaVidal2002}
F. J. Garc\'ia-Vidal and L. Mart\'in-Moreno,
``Transmission and focusing of light in one-dimensional periodically nanostructured metals,''
Phys. Rev. B \textbf{66}, 155412 (2002).

\bibitem{Novotny2012}
L. Novotny and N. van Hulst,
``Antennas for light,''
Nat. Photonics \textbf{5}, 83--90 (2011).

\bibitem{Xie2004}
Y. Xie, A. R. Zakharian, J. V. Moloney, and M. Mansuripur,
``Transmission of light through slit apertures in metallic films,''
Opt. Express \textbf{12}, 6106--6121 (2004).

\bibitem{Han2021}
Y. Han, Y. Lin, W. Ma, J. G. Korvink, H. Duan, and Y. Deng,
``Nanoantennas inversely designed to couple free space and a metal-insulator-metal waveguide,''
Nanomaterials \textbf{11}, 3219 (2021).

\end{thebibliography}
\end{document}